\documentclass[aps, prb, twocolumn, showpacs, superscriptaddress]{revtex4}
\usepackage{amsmath, amsfonts, amssymb}
\usepackage{bm}
\usepackage{hyperref, verbatim}
\usepackage{graphicx}

\begin{document}

\def\Re {\mbox{Re}}
\def\Im {\mbox{Im}}
\newcommand{\avg}[1]{\langle#1\rangle}
\newcommand{\odiff}[2]{\frac{\di #1}{\di #2}}
\newcommand{\pdiff}[2]{\frac{\partial #1}{\partial #2}}
\newcommand{\di}{\mathrm{d}}
\newcommand{\ii}{i}
\newcommand{\norm}[1]{\left\| #1 \right\|}
\renewcommand{\vec}[1]{\mathbf{#1}}
\newcommand{\ket}[1]{|#1\rangle}
\newcommand{\bra}[1]{\langle#1|}
\newcommand{\pd}[2]{\langle#1|#2\rangle}
\newcommand{\tpd}[3]{\langle#1|#2|#3\rangle}
\renewcommand{\vr}{{\vec{r}}}
\newcommand{\vk}{{\mathbf{k}}}
\renewcommand{\ol}[1]{\overline{#1}}
\newcommand{\comments}[1]{}

\title{Topological Protection of Majorana Qubits}

\author{Meng Cheng}
\affiliation{Station Q, Microsoft Research, Santa Barbara, CA 93106-6105}

\affiliation{Condensed Matter Theory Center, Department of Physics,
University of Maryland, College Park, Maryland 20742, USA }

\affiliation{
Kavli Institute for Theoretical Physics, University of California, Santa Barbara, CA, 93106}

\author{Roman M.~Lutchyn}
\affiliation{Station Q, Microsoft Research, Santa Barbara, CA 93106-6105}
\date{\today}

\author{S. Das Sarma}
\affiliation{Condensed Matter Theory Center, Department of Physics,
University of Maryland, College Park, Maryland 20742, USA }

\date{\today}

\begin{abstract}
	We study the stability of the topological quantum computation proposals involving  Majorana fermions against thermal fluctuations. We use a minimal realistic model of a spinless $p_x+ip_y$ superconductor and consider effect of excited midgap states localized in the vortex core as well as of transitions above the bulk superconducting gap on the quasiparticle braiding, interferometry-based qubit read-out schemes, and quantum coherence of the topological qubits. We find that thermal occupation of the midgap states does not affect adiabatic braiding operations but leads to a reduction in the visibility of the interferometry measurements. We also consider quantum decoherence of topological qubits at finite temperatures and calculate their decay rate which is associated with the change of the fermion parity and, as such, is exponentially suppressed at temperatures well below the bulk excitation gap. Our conclusion is that the Majorana-based topological quantum computing schemes are indeed protected by the virtue of the quantum non-locality of the stored information and the presence of the bulk superconducting gap.
\end{abstract}
%03.67.Lx  Quantum computation
%03.65.Vf  Topological phases (quantum mechanics
%74.20.Rp 	Pairing symmetries (other than s-wave)
%05.30.Pr	Fractional statistics systems (anyons, etc.)
%71.10.Fd 	Lattice fermion models (Hubbard model, etc.)
%03.67.Pp   Quantum error correction and other methods for protection against decoherence (see also 03.65.Yz Decoherence; open systems; quantum statistical methods; for decoherence in Bose-Einstein condensates, see 03.75.Gg)
%74.90.+n	Other topics in superconductivity (restricted to new topics in section 74)
%67.10.Db 	Fermion degeneracy
\maketitle

\section{Introduction}

Topological quantum computation, based on the encoding of quantum information in non-local degrees of freedom, provides a promising route to fight quantum decoherence~\cite{Kitaev_AP03, nayak_RevModPhys'08, blueprint}. Due to the presence of environmental interactions, decoherence is the foremost
challenge in any conventional quantum computation schemes since quantum error correction protocols typically have very severe constraints on the amount of error that can be corrected in a fault-tolerant manner. Topological quantum computation (TQC) utilizes topological degeneracy of certain low-dimensional systems believed to host non-Abelian quasiparticles (anyons): fractional quantum Hall states ({\it e.g.} at filling factor $\nu=5/2$)~\cite{Moore_NPB91, Nayak_NPB96, dassarma_prl'05}, certain exotic lattice spin systems~\cite{Kitaev_AP03}, and topological superconductors~\cite{read_prb'00, Kitaev_Majorana, Ivanov_PRL'01}. The latter has received tremendous attention recently~\cite{Wilczek_NatPhys, Stern_Nat, Franz_popular, Levi_PhysToday} and a number of possible candidates for topological superconductivity has been proposed including strontium ruthenate~\cite{DasSarma_PRB'06}, topological insulator-superconductor heterostructure~\cite{Fu_PRL08, Cook_PRB2011}, semiconductor-superconductor heterostructure~\cite{Sau_PRL10, Alicea_PRB10, Lutchyn_PRL2011, Oreg_PRL2010, Qi_PRB2010} and non-centrosymmetric superconductors~\cite{Sato_PRB09}. In all these systems, non-Abelian Ising anyons are realized as Majorana zero-energy modes bound to certain topological defects and obey non-Abelian braiding statistics~\cite{Moore_NPB91,Nayak_NPB96, Bonderson_PRB2011, Ivanov_PRL'01, Alicea_NatPhys2011}.
Given that Majorana zero-energy quasiparticles are described by hermitian operators $\gamma=\gamma^\dag$, one can show by constructing a non-local Dirac operator $\hat{c}$ out of two spatially-separated Majorana operators $\hat{\gamma}_{i=1,2}$ ({\it i.e.} $\hat{c}=\hat{\gamma}_1+i\hat{\gamma}_2$) that there are $2^{n-1}$ degenerate states for a given overall fermion parity with $2n$ Majorana zero-energy modes at fixed positions. The information is encoded in the occupied or unoccupied states of the non-local Dirac fermion modes. This is a crucial concept for the Majorana-based TQC proposals. As long as global fermion parity in the system is preserved, one can design fault-tolerant quantum computation schemes at sufficiently low temperatures.

In this paper we investigate the effect of finite-temperature thermal fluctuations on  three key aspects of topological quantum computation: quantum coherence of the topological qubits, topologically-protected quantum gates and the read-out of qubits. Since the information is encoded in non-local degrees of freedom of the ground state many-body wavefunction, it is important to keep the system close to the ground state. However, any systems realized in the laboratory are operated at a finite temperature $T>0$. To prevent uncontrollable thermal excitations, it is generally accepted that $T$ has to be way below the bulk excitation gap. However, complications appear when there exist various types of single-particle excitations with different magnitudes of gaps which can change the occupation of the non-local fermionic modes. Note that throughout the paper we assume that Majorana fermions are sufficiently far away from each other and neglect exponentially small energy splitting due to inter-vortex tunneling. The effect of these processes on topological quantum computing has been discussed elsewhere~\cite{Cheng_PRL09, Cheng_PRB2010b}. Another trivial effect not considered in this work is a situation where the fermion parity conservation is explicitly broken by the Majorana mode being in direct contact with a bath of fermions (electrons and holes) where obviously the Majorana will decay into the fermion bath, and consequently decohere.  Such situations arise, for example, in current topological insulators where the existence of the bulk carriers (invariably present due to the unintentional bulk doping) would make any surface non-Abelian Majorana mode disappear rather rapidly.  Another situation that has recently been considered in this context~\cite{Sau_periodic} is the end Majorana mode in a one-dimensional nanowire being in contact with the electrons in the non-superconducting part of the semiconductor, leading to a zero-energy Majorana resonance rather than a non-Abelian Majorana bound state at zero energy.  The fact that the direct coupling of Majorana modes to an ordinary fermionic bath will lead to its decoherence is rather obvious and well-known, and does not require a general discussion since such situations must be discussed on a case by case basis taking into account the details of the experimental systems.  In particular, the reason the quantum braiding operations in Majorana-based systems involves interferometry is to preserve the fermion parity conservation. Our theory in the current work considers the general question of how thermal fluctuations at finite temperatures affect the non-Abelian and the non-local nature of the Majorana mode.

We consider a simple model for two-dimensional chiral $p_x+ip_y$ superconductor where Majorana zero-energy states are hosted by Abrikosov vortices. The quasiparticle excitations in this system are divided into two categories: a) Caroli-de Gennes-Matricon (CdGM) or so-called midgap states localized in the vortex core with energies below the bulk superconducting gap~\cite{Caroli_PL'64, Kopnin_PRB'91} (the gap that separates the zero-energy state to the lowest CdGM state is called the mini-gap $\Delta_M$);  b) extended states with energies above the bulk quasiparticle gap which is denoted by $\Delta$. The natural question arising in this context is how these two types of excitations affect topological quantum computation using the Majorana zero-energy states at finite temperature. This question is very relevant in the context of strontium ruthenate as well as other weak-coupling BCS superconductors where the Fermi energy $E_F$ is much larger than the superconducting gap  $\Delta$ in which case $\Delta_M\propto \Delta^2/E_F \ll \Delta$. We mention in passing here that the semiconductor-based Majorana proposals~\cite{Sau_PRL10,Alicea_PRB10, Lutchyn_PRL2011, Oreg_PRL2010} do not have low-lying CdGM states because the minigap $\Delta_M\sim \Delta$~\cite{robustness} due to the small Fermi energy in the semiconductor. If the temperature is substantially below the minigap, i.e. $T\ll \Delta_M$, obviously all excited states can be safely ignored. However, such low temperatures with $T\ll \Delta_M$ can be hard to achieve in the laboratory since for typical superconductors $\Delta/E_F\sim 10^{-3}-10^{-4}$. We note that even in the semiconductor sandwich structures the Majorana energetics obey the inequality $\Delta_M<\Delta$ since in general $E_F>\Delta$ even in the semiconductor-based systems in view of the fact that typically $\Delta\sim 1\,\text{K}$.  This makes our consideration in this paper of relevance also to the semiconductor-based topological quantum computing platforms. In this paper we investigate the non-trivial intermediate temperature regime $\Delta_M \ll T < \Delta$. To make this paper more pedagogical, we will use a simple physical model that captures the relevant physics. We find that the presence of the excited midgap states localized in the vortex core does not effect braiding operations. However, the midgap states do affect the outcome of the interferometry experiments.

We also study the quantum dynamical evolution and obtain equations of motion for the reduced density matrix assuming that the finite temperature is set by a bosonic bath (e.g. phonons). We find that the qubit decay rate $\lambda$ is given by the rate of changing fermion parity in the system and is exponentially suppressed (i.e. $\lambda \propto \exp(-\Delta/T)$) at low temperatures in a fully-gapped $p_x+ip_y$ superconductor. In this context, we make some comments about Refs.[\onlinecite{Goldstein_wrong}] claiming to obtain different results regarding the effect of thermal fluctuations.

The paper is organized as follows. In Sec.~\ref{sec:braiding} we generalize the notion of non-Abelian braiding to a finite temperature and show that braiding is not affected by CdGM bound states. In Sec.~\ref{sec:interferometry} we show that the midgap states are important for interferometry experiments and generally reduce the visibility of the signal. In Sec.~\ref{sec:decoherence} we study the problem of qubit decoherence and effects of thermal fluctuations. Finally, we conclude in Sec.\ref{sec:conclusion}. Some technical details are given in the Appendices.

\section{Non-Abelian Braiding in the Presence of Midgap States}\label{sec:braiding}
In this section we address the question of how the midgap states affect the non-Abelian statistics at finite temperature. The usual formulation of the non-Abelian statistics as unitary transformation of the ground states does not apply, since at finite temperature the system has to be described as a mixed state. We need to generalize the notion of the non-Abelian braiding in terms of physical observables~\cite{Akhmerov_PRB2010}. This can be done as the following: consider a topological qubit made up by four vortices labeled by $a=1,2,3,4$. Each of them carries a Majorana zero-energy state, whose corresponding quasiparticle is denoted by $\hat{\gamma}_{a0}$ which satisfies $\hat{\gamma}_{a0}^2=1, \hat{\gamma}_{a0}=\hat{\gamma}_{a0}^\dag$. There are other midgap states in the vortex core which are denoted by $\hat{d}_{ai},i=1,2,\dots, m$. (Actually the number of midgap states is huge and the midgap spectrum eventually merges with the bulk excitation spectrum. However, since we are interested in $T\ll \Delta$, we can choose an energy cutoff $\Lambda$ such that $T\ll \Lambda \ll \Delta$ and only include those midgap states that are below $\Lambda$.) It is convenient to write $\hat{d}_{ai}=\hat{\gamma}_{a,2i-1}+i\hat{\gamma}_{a,2i}$, so each vortex core carries odd number of Majorana fermions $\hat{\gamma}_{ai}, i=0,1,\dots, 2m$.

Having the notations set up, we now define a generalized Majorana operator $\hat{\Gamma}_a=i^m\prod_{i=0}^{2m}\hat{\gamma}_{ai}$. It is straightforward to check that $\{\hat{\Gamma}_a,\hat{\Gamma}_b\}=2\delta_{ab}$. We then define the fermion parity shared by a pair of vortices $
\hat{\Sigma}_{ab}=i\hat{\Gamma}_a\hat{\Gamma}_b$.
The topological qubit can be uniquely specified by a set of measurements of the expectation value of the following Pauli matrices $\hat{\bm{\sigma}}=(\hat{\sigma}_x,\hat{\sigma}_y,\hat{\sigma}_z)$:
\begin{equation}
	 \hat{\sigma}_x=\hat{\Sigma}_{32},\hat{\sigma}_y=\hat{\Sigma}_{13},\hat{\sigma}_z=\hat{\Sigma}_{21}.
	\label{}
\end{equation}
The non-Abelian braiding can be represented as the transformation of $\langle \hat{\bm{\sigma}}\rangle$.

Now we list the key assumptions to establish the non-Abelian properties of the vortices:
\begin{enumerate}
	\item The fermion parity $\hat{\Sigma}_{ab}$ is a physical observable that can be measured by suitable interferometry experiments, even at finite temperature.
	\item All the bound states remain localized together with the zero-energy state when the vortices are transported. Therefore they can be considered as one composite system.
	\item The tunneling processes of fermions between different vortices and transitions to the gapped continuum are exponentially suppressed due to the presence of the bulk superconducting gap.
\end{enumerate}
Under these conditions, the only local dynamical processes are the transitions of fermions between the localized bound states, e.g. scattering by collective excitations like phonons. However, such processes necessarily conserve $\hat{\Gamma}_a$, therefore also the parities $\hat{\Sigma}_{ab}$.

To see this explicitly, the state of the qubit is described by the density matrix $\hat{\rho}(t)$. Because we are truncating the whole Hilbert space to include only those below our cutoff $\Lambda$, it is necessary to use the time-dependent instantaneous basis~\cite{Cheng_PRB2011}. We can generally consider transitions between the various states induced by four-fermion interactions or coupling to a bosonic bath. To be specific, we write down the Hamiltonian of the system:
\begin{equation}
	\hat{H}=\hat{H}_0+\hat{H}_\text{int}.
	\label{}
\end{equation}
Here $\hat{H}_0$ is the Hamiltonian of the BCS superconductor with vortices, whose positions $\vec{R}_i$ are time-dependent. At each moment of time $\hat{H}_0$ can be diagonalized, yielding a set of complete eigenbasis which are represented by the time-dependent generalization of the aforementioned Bogoliubov quasiparticles $\hat{\gamma}_{a0}(t), \hat{d}_{ai}(t)$. $\hat{H}_\text{int}$ describes all kinds of perturbations that are allowed under the assumptions.

Without going into the details of microscopic calculations, we write down the general Lindblad form of the master equation~\cite{lindblad} governing the time-evolution of the density matrix:
\begin{equation}
	 \odiff{\hat{\rho}}{t}=\pdiff{\hat{\rho}}{t}-i[\hat{H}_t(t),\hat{\rho}]+\hat{S}\hat{\rho} \hat{S}^\dag-\frac{1}{2}\{\hat{S}^\dag \hat{S},\hat{\rho}\}.
	\label{}
\end{equation}
The $\pdiff{\hat{\rho}}{t}$ denotes the change of $\hat{\rho}$ solely due to the change of basis states. Here $\hat{H}_t$ describes the (effective) unitary evolution of the density matrix due to transitions between different fermionic states and the Lindblad superoperators $\hat{S}$ corresponds to non-unitary evolution induced by system-environment coupling. Our assumption on the locality of the interactions in the system implies that
\begin{equation}
	[\hat{H}_t, \hat{\Sigma}_{ab}]=0, [\hat{S},\hat{\Sigma}_{ab}]=0.
	\label{eqn:commute}
\end{equation}

The time evolution of the expectation values of $\bm{\sigma}(t)$ is given by
\begin{equation}
	 \odiff{\langle\bm{\sigma}\rangle}{t}=\odiff{}{t}\text{Tr}\,\bm{\sigma}\hat{\rho}=\text{Tr}\,\pdiff{\bm{\sigma}}{t}\hat{\rho}(t)+\text{Tr}\,\bm{\sigma}\odiff{\hat{\rho}}{t}.
	\label{}
\end{equation}

With \eqref{eqn:commute}, it is straightforward to check that
\begin{equation}
	 \text{Tr}\,\bm{\sigma}[\hat{H},\hat{\rho}]=0,\text{Tr}\,\bm{\sigma}\big(\hat{S}\hat{\rho} \hat{S}^\dag-\frac{1}{2}\{\hat{S}^\dag \hat{S},\hat{\rho}\}\big)=0.
	\label{}
\end{equation}
Therefore we have
\begin{equation}
	 \odiff{\langle\bm{\sigma}(t)\rangle}{t}=\text{Tr}\,\pdiff{\bm{\sigma}(t)}{t}\hat{\rho}(t)+\text{Tr}\,\bm{\sigma}(t)\pdiff{\hat{\rho}(t)}{t}=\partial_t\text{Tr}\,{[\bm{\sigma}(t)\hat{\rho}(t)]}.
	\label{eqn:braid1}
\end{equation}
As we have defined, $\partial_t$ means that all changes come from the change in the basis $\{\hat{\gamma}_{ai}(t)\}$. Since after the braiding the system returns to its initial configuration, the operators $\hat{\gamma}_{ia}$ undergo unitary transformations. So if the braiding starts at $t=t_i$ and ends at $t=t_f$, we have the simple result $\langle\bm{\sigma}(t_i)\rangle=\langle\bm{\sigma}(t_f)\rangle$. However, the operators $\hat{\Gamma}(t_f)$ are different from $\hat{\Gamma}(t_i)$.
One can easily verify that the operators $\hat{\Gamma}_a$ satisfy Ivanov's rule~\cite{Ivanov_PRL'01, Akhmerov_PRB2010} under braiding of vortices $a$ and $b$:
\begin{equation}
	\hat{\Gamma}_a\rightarrow \hat{\Gamma}_b, \hat{\Gamma}_b\rightarrow -\hat{\Gamma}_a.
	\label{}
\end{equation}
And the transformation of $\langle \hat{\bm{\sigma}}\rangle$ is identical to the case without any midgap states. In conclusion, in terms of physically measurable quantities, the non-Abelian statistics is well-defined in the presence of excited midgap states localized in the vortex core.

This result is thoroughly non-obvious because it may appear on first sight that arbitrary thermal occupancies of the mid-gap excited states would completely suppress the non-Abelian nature of the system since the Majorana mode resides entirely at zero energy and not in the excited mid-gap states.

\section{Interferometry in the Presence of Midgap States}\label{sec:interferometry}
We now discuss the effect of the midgap states in interferometry experiments designed for the qubit read-out~\cite{SDS_PRL2005, Stern_PRL2006, Bonderson_PRL2006, Bonderson_AnnPhys2008} There is a number of recent proposals for interferometry experiments in topological superconductors~\cite{Akhmerov_int'09, Sau_arxiv1004, Grosfeld_PRB2010}.
%\begin{comment}
\begin{figure}
	\begin{center}
		\includegraphics[width=\columnwidth]{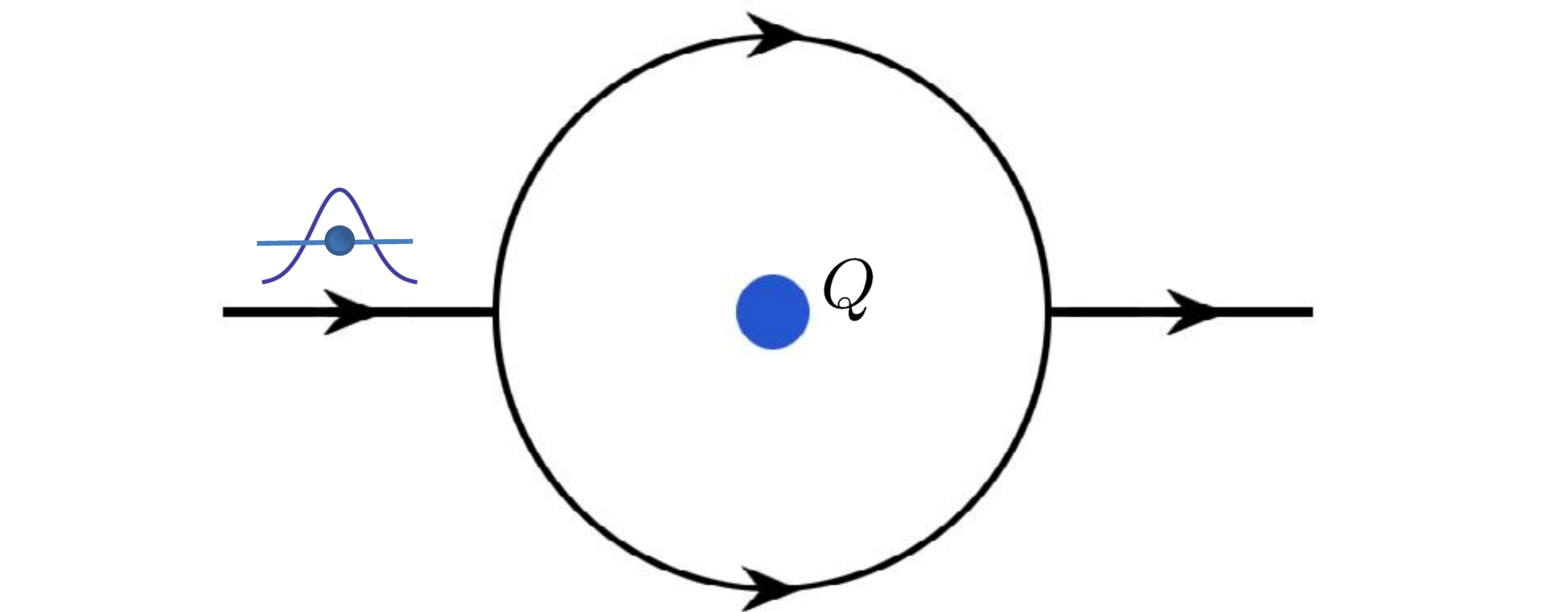}
	\end{center}
	\caption{Mach-Zehnder interferometer proposed in Ref. [\onlinecite{Grosfeld_PNAS}] for topological qubit detection. Due to the Aharonov-Casher effect, the vortex current is sensitive to the charge enclosed. Long Josephson junction between two topological superconductors carries allows for Josephson vortices (fluxons) that carry Majorana zero-energy modes.}
	\label{fig:ac}
\end{figure}
%\end{comment}
In this paper, we use an example of the Mach-Zehnder interferometer proposed by Grosfeld and Stern~\cite{Grosfeld_PNAS} based on Aharonov-Casher (AC) effect. In this proposal, a Josephson vortex (fluxon) is driven by supercurrent to circumvent a superconducting island. The fluxon appearing at the interface between two topological $p_x+ip_y$ superconductors carries a zero-energy Majorana modes, and behaves as a non-Abelian anyon. Therefore, the vortex current around the central superconductor is sensitive to the topological content of the enclosed superfluid. (We refer the reader to Ref. [\onlinecite{Grosfeld_PNAS}] for more details.) Indeed, the vortex current is proportional to the total tunneling amplitude:
\begin{equation}
	\begin{split}
		J_v&\propto |(t_L \hat{U}_L+t_L \hat{U}_R)\ket{\Psi_0}|^2\\
		&=|t_L|^2+|t_R|^2+2 \text{Re}\{t_L^*t_R\tpd{\Psi_0}{\hat{U}_L^{-1}\hat{U}_R}{\Psi_0}\}\\
		&=|t_L|^2+|t_R|^2+2 \text{Re}\{t_L^*t_R e^{i\varphi_\text{AC}}\tpd{\Psi_0}{\hat{M}}{\Psi_0}\}.
\end{split}
	\label{}
\end{equation}
Here $\ket{\Psi_0}$ is the initial state of the system and $\hat{U}_L$ and $\hat{U}_R$ are the unitary evolution operators for the fluxon taking  the two respective paths. $\varphi_\text{AC}$ is the Aharonov-Casher phase accumulated by the fluxon: $\varphi_\text{AC}=\pi Q/e$. Here $Q$ is the total charge enclosed by the trajectory of the fluxon, including the offset charge $Q_\text{ext}$ set by external gate and the fermion parity $n_p$ of the low-energy fermionic states:
\begin{equation}
	Q=Q_\text{ext}+e n_p.
	\label{}
\end{equation}
$\hat{M}$ encodes the transformation solely due to the braiding statistics of the non-Abelian fluxon around $n$ non-Abelian vortices.

If the superconducting island contains no vortices, then $\hat{M}=1$ and the interference term is solely determined by the AC phase. The magnitude of the vortex current shows an oscillation:
\begin{equation}
	J_v=J_{v0}\Big[1+\zeta\cos\Big(\frac{\pi Q}{e}\Big)\Big].
	\label{}
\end{equation}
Here $\zeta$ is the visibility of the interference.

When $n$ is odd, there is no interference because $\hat{M}\ket{\Psi_0}$ and $\ket{\Psi_0}$ have different fermion parity, implying $\tpd{\Psi_0}{\hat{M}}{\Psi_0}=0$.  The vortex current becomes independent of the charge encircled. Therefore, the disappearance of the interference can be used as a signature of the non-Abelian statistics of the vortices.

We now consider a situation where the non-Abelian fluxon has midgap states other than the Majorana bound state. The internal state of the fluxon then also depends on the occupation of these midgap states. As we have argued in the previous section, as far as braiding is concerned the non-Abelian character is not affected at all by the presence of midgap states. So the interference still vanishes when there are odd numbers of non-Abelian vortices in the island. On the other hand, when there are no vortices in the island, transitions to the midgap states can significantly reduce the visibility of the interference term $\zeta$.

To understand quantitatively how the visibility of the interference pattern is affected by the midgap state, let us consider the following model of the fluxon. Since we are interested in the effect of midgap states, we assume there is only one midgap state and model the probe vortex by a two-level system, or spin $1/2$, with the Hilbert space $\{\ket{0},\ket{1}\}$. Here $\ket{1}$ denotes the state with the midgap state occupied. We also assume that the charge enclosed by the interference trajectory $Q=0$ so we can neglect the AC phase. The Hamiltonian is then given by
\begin{equation}
	\hat{H}=\ket{L}\bra{L} \otimes \hat{H}_L+\ket{R}\bra{R}\otimes \hat{H}_R.
	\label{}
\end{equation}
where $\hat{H}_{L,R}$ is given by
\begin{equation}
	\hat{H}_{\eta}=\frac{\Delta}{2} \sigma_z+\sigma_x\sum_{k}g_k(\hat{a}_{\eta,k}^\dag + \hat{a}_{\eta,k})+\sum_k\omega_{\eta,k} \hat{a}_{\eta,k}^\dag \hat{a}_{\eta,k}.
	\label{}
\end{equation}
Here $\eta=L,R$. Notice that we assume the bath couples to the fluxon locally so we introduce two independent baths for $L$ and $R$ paths. The unitary evolution at time $t$ is then factorizable:
\begin{equation}
	\hat{U}(t)=\ket{L}\bra{L}\otimes \hat{U}_L(t)+\ket{R}\bra{R}\otimes \hat{U}_R(t).
	\label{}
\end{equation}
Given initial state $\hat{\rho}(0)=\hat{\rho}_{\text{path}}\otimes\hat{\rho}_{\text{s}}\otimes\hat{\rho}_\text{bath}$, we can find the off-diagonal component of the final state $\hat{\rho}(t)=\hat{U}(t)\hat{\rho}(0)\hat{U}^\dag(t)$, corresponding to the interference, as
\begin{equation}
	\begin{split}
		 \lambda_{LR}&=\mathrm{Tr}\,\left[\hat{U}_L(t)\hat{U}^\dag_R(t)\hat{\rho}_\text{s}\otimes\hat{\rho}_\text{bath}\right]\\
	 &=\mathrm{Tr}\Big[\rho_\text{s}\mathrm{Tr}_{L}[\hat{\rho}_{\text{bath},L}\hat{U}_L(t)] \mathrm{Tr}_{R}[\hat{\rho}_{\text{bath},R}\hat{U}_R(t)]\Big].
\end{split}
	\label{}
\end{equation}

Now we evaluate $\hat{W}_\eta(t)=\mathrm{Tr}_{\eta}[\hat{\rho}_{\text{bath},\eta}\hat{U}_\eta(t)]$ (notice $\hat{W}_\eta$ is still an operator in the spin Hilbert space). We drop the $\eta$ index in this calculation. First we switch to interaction picture and the evolution operator $\hat{U}(t)$ can be represented formally as $\hat{U}(t)=\mathcal{T} \exp\{-i\int_0^t \di t'\,\hat{H}_1(t')\}$ where
\begin{equation}
	\hat{H}_1(t)=\sum_k g_k (\sigma^+e^{i\Delta t/2}+\sigma^- e^{-i\Delta t/2})(\hat{a}_k^\dag e^{i\omega_k t}+\hat{a}_k e^{-i\omega_k t}).
	\label{}
\end{equation}
Following the derivation of the master equation for the density matrix, we can derive a ``master equation'' for $\hat{W}(t)$ under the Born-Markovian approximation:
\begin{equation}
	\odiff{\hat{W}}{t}=-\gamma(\ol{n}+1/2+\sigma_z/2)\hat{W},
	\label{}
\end{equation}
where $\gamma=\pi\sum_k g_k^2\delta(\omega_k-\Delta), \ol{n}=\gamma^{-1}\sum_k g_k^2 \ol{n}_k\delta(\omega_k-\Delta)$.

Therefore, the visibility of the interference, proportional to the trace of $\hat{W}$, is given by
\begin{equation}
	\zeta\propto \mathrm{Tr}[\hat{W}(t)\rho_\text{s}]\propto e^{-\gamma \ol{n} t}=e^{-\gamma\ol{n}L/v}.
	\label{}
\end{equation}

We notice that the model we have used is of course a simplification of the real fluxon. We only focus on the decoherence due to the midgap states and  assume that only one such state is present. In reality, there could be many midgap states which lead to a stronger suppression of visibility.
%Assuming the bath is in its thermal equilibrium with temperature $T$ and at $t=0$ the density matrix is factorizable, we evaluate $W(t)$ perturbatively to the second order in system-bath couplings:
%\begin{equation}
%	\begin{split}
%	|W(t)|\approx& 1-\frac{1}{2}\int_0^\infty \di \omega J(\omega)\ol{n}(\omega)\Big[\frac{\sin(\omega-\varepsilon)t/2}{(\omega-\varepsilon)/2}\Big]^2\\
%	\approx& 1-\pi tJ(\varepsilon)\ol{n}(\epsilon)\approx e^{-\pi tJ(\varepsilon)\ol{n}(\epsilon)}
%\end{split}
%	\label{}
%\end{equation}
%Here $J(\omega)$ is the spectra density of the bath defined as
%	$J(\omega)=\sum_k g_k^2\delta(\omega-\omega_k)$.
%And $\ol{n}(\omega)=1/(e^{\omega/T}-1)$ is the boson distribution function.
%The last line is valid when we consider time scale $t\gg \varepsilon^{-1}$.
\comments{
So far we have been focusing on the midgap states in the fluxon, i.e. the vortex that is used to probe the Majorana fermion in the interferometer.  The midgap bound states carry charge which can be calculated using the charge operator $\hat{Q}=e\int\di^2\vr:\hat{\psi}^\dag(\vr)\hat{\psi}(\vr):$. Then the charge of a bound state with wavefunction $(u_n, v_n)$ is given by
\begin{equation}
	Q_n=\int\di^2\vr(|u_n(\vr)|^2-|v_n(\vr)|^2).	
	\label{}
\end{equation}
If the bound state is not Majorana ($E_n\neq 0$), then generically $Q_n\neq 0$. For spinless $p_x+ip_y$ BCS superconductor, we estimate the charge of the first excited bound state, assuming $k_F\xi\gg 1$:
\begin{equation}
	Q\approx
	\begin{cases}
		-e\frac{\ln k_F\xi}{k_F\xi} & k_F\xi \gg 1\\
		-e\big[1-\frac{15}{4}(k_F\xi)^4 \big]& k_F\xi \ll 1
	\end{cases}
	.
	\label{}
\end{equation}
Therefore $|Q|/e\ll 1$ for weak-coupling superconductors where $k_F\xi\gg 1$. For topological superconductors realized in spin-orbit-coupled semiconductors, the Fermi momentum $k_F$ is smaller so one might expect a larger charge, but there the energy of the midgap states are of the same order as the bulk gap and the thermal occupation of them can be neglected.
}

The above interferometor is able to detect the existence of non-Abelian vortices which requires that the Josephson vortex (i.e. fluxon) also has Majorana midgap states. To fully read out a topological qubit, one needs to measure the fermion parity of the qubit. This can also be done using interferometry experiments with flux qubits, essentially making use of the AC effect of Josephson vortices~\cite{Sau_PRA2010, Hassler_NJP2010}.

Another relevant question is whether the thermal excitations of the (non-Majorana) midgap states localized in the vortex core have any effects on the interferometry. Since the interferometry is based on AC effect where vortex acquires a geometric phase after circling around some charges, one might naively expect that the interferometric current might depend on the occupation of the midgap states due to the charge associated with the midgap states (i.e. $Q_n=e\int\di\vr\,(|u_n|^2-|v_n|^2)$). The situation is more subtle, however, once one takes into account the screening effect due to the superfluid condensate. The kinetics of the screening process is beyond the scope of this paper. However, assuming equilibrium situation more careful inspection discussed in the Appendix~\ref{appendixB} shows the geometric phases acquired by the Josephson vortices only depends on the total fermion parity in the low-energy midgap states (even if they are not Majorana zero-energy modes) and the offset charge set by the external gate voltage.

\section{Depolarization of Qubits at Finite Temperature}\label{sec:decoherence}
We now study the coherence of the topological qubit itself. From our discussion on the effect of bound states in the vortex core, it is clear that decoherence only occurs when the qubit is interacting with a macroscopically large number of fermionic degrees of freedom, a fermionic bath. An example of such a bath is provided by the continuum of the gapped quasiparticles in the topological superconductor. Once the Majorana fermion is coupled to the bath via a tunneling Hamiltonian, the fermion occupation in the qubit can leak into the environment, resulting in the depolarization of the qubit. It is then crucial to have a fully gapped quasiparticle spectrum to ensure that such decoherence is exponentially small, as will be shown below.

To study the decay of a Majorana zero mode, we consider two such modes, $\hat{\gamma}_1$ and $\hat{\gamma}_2$, forming an ordinary fermion $c=\hat{\gamma}_1+i\hat{\gamma}_2$. The gapped fermions are coupled locally to $\hat{\gamma}_1$, without any loss of generality. The coupling is mediated by a bosonic bath. The Hamiltonian then reads
\begin{equation}
	\begin{split}
		\hat{H}=i\varepsilon\hat{\gamma}_1\hat{\gamma}_2+\sum_k\epsilon_k \hat{d}_k^\dag \hat{d}_k+\sum_l\omega_l\hat{a}^\dag_l \hat{a}_l\\
		+i\sum_{kl} g_{kl}\hat{\gamma}_1(\hat{d}_k^\dag+\hat{d}_k)(\hat{a}_l^\dag + \hat{a}_l).
\end{split}
	\label{}
\end{equation}
Here $\hat{d}_k$ is the annihilation operator of the gapped fermions with quantum number $k$ and energy $\varepsilon_k$. $\hat{a}_l$ is the annihilation operators of the bosonic bath.

The density matrix of the system evolves according to the equation of motion $\dot{\hat{\rho}}=-i[\hat{H},\hat{\rho}]$. Since we are interested in the qubit only, we will derive the master equation for the reduced density matrix $\hat{\rho}_\text{r}$, tracing out the bosonic bath and the gapped fermions.
\begin{equation}
	 \odiff{\hat{\rho}_\text{r}}{t}=-\lambda\big[\hat{\rho}_\text{r}-\hat{\gamma}_1(-1)^{\hat{n}}\hat{\rho}_\text{r}(-1)^{\hat{n}}\hat{\gamma}_1\big],
	\label{eqn:master}
\end{equation}
where
\begin{equation}
	 \lambda=2\sum_{kl}g_{kl}^2\big[(1-n^f_{k})n^b_{l}+n^f_{k}(n^b_{l}+1)]\delta(\varepsilon_k-\omega_l).
	\label{}
\end{equation}
Here $n^{f}_k=1/(e^{\varepsilon_k/T}+1), n^b_l=1/(e^{\omega_l/T}-1)$ are the Fermi and Bose distribution functions.
The derivation of \eqref{eqn:master} is presented in the Appendix~\ref{appendixA}. Notice that at low temperatures $T \ll \Delta$, due to energy conservation, both $n^b_{l}$ and $n^f_{k}$ are suppressed by the Gibbs factor $e^{-\Delta/T}$. Therefore, the rate $\lambda\sim e^{-\Delta/T}$.

Then the polarization of the qubit $\langle \sigma_z\rangle=\text{Tr}[\sigma_z\hat{\rho}_\text{r}]$ satisfies $d_t\langle\sigma_z\rangle=-2\lambda \langle\sigma_z\rangle$. Therefore the lifetime of the topological qubit is given by $T_1\sim \lambda^{-1}$. Physically, this is reasonable since we introduce tunneling term between the Majorana fermion and the gapped fermionic environment so the fermion parity of the qubit is no longer conserved. It is expected that $\lambda$ is determined by the exponential factor $e^{-\Delta/T}$ when $T\ll \Delta$. Therefore, this provides a quantitative calibration of the protection of the topological qubit at finite temperature. In the high-temperature limit $T\gg \Delta$, the distribution function scales linearly with $T$ so the decay rate is proportional to $T$. This is quite expected since $T\gg \Delta$, the gap does not play a role. We note that a recent work by Goldstein and Chamon~\cite{Goldstein_wrong} studying the decay rate of Majorana zero modes coupled to classical noise essentially corresponds to the high-temperature limit of our calculation $T\gg \Delta$ and, as such, does not apply to any realistic system where the temperature is assumed to be low, i.e. $T\ll \Delta$.In fact, in the trivial limit of $\Delta \ll T$, the Majorana decoherence is large and weakly temperature dependent because the fermion parity is no longer preserved and the fermions can simply leak into the fermionic bath~\cite{Clarke_NJP2011}.  By definition, this classical limit of $T\gg\Delta$ is of no interest for the topological quantum computation schemes since the topological superconductivity itself (or for that matter, any kind of superconductivity) will be completely absent in this regime. Our result makes sense from the qualitative considerations: quantum information is encoded in non-local fermionic modes and changing fermion parity requires having large thermal fluctuations or external noise sources with finite spectral weight at frequencies $\omega\sim\Delta$.
Furthermore, it is important to notice that such relaxation can only occur when the qubit is coupled to a continuum of fermionic states  which renders the fermion parity of the qubit undefined. Intuitively, the fermion staying in the qubit can tunnel to the continuum irreversibly, which is accounted for by the procedure of ``tracing out the bath'' in our derivation of the master equation. It is instructive to compare this result to a different scenario, where the zero-energy fermionic state is coupled to a fermionic state (or a finite number of them) instead of a continuum. In that case, due to hybridization between the states the fermion number oscillates between the two levels with a period (recurrence time) determined by the energy difference $\Delta E$ between them. The expectation value of the fermion number (or spectral weight) in the zero-energy state is depleted and oscillatory in time, but will not decay to zero.

The above derivation can be straightforwardly generalized to $N>2$ Majorana fermions, each coupled locally to gapped fermions and bosonic bath.
\begin{equation}
	 \odiff{\hat{\rho}_\text{r}}{t}=-\sum_{i=1}^N\lambda_i\big[\hat{\rho}_\text{r}-\hat{\gamma}_i(-1)^{\hat{n}_i}\hat{\rho}_\text{r}(-1)^{\hat{n}_i}\hat{\gamma}_i\big],
	\label{eqn:master2}
\end{equation}
The depolarization of the qubit can be calculated in the same fashion.

%One may wonder that the calculation above makes no use of the fact that the qubit is a non-local object, so it can be stated as well to an ordinary fermionic bound state. However, we would like to emphasize here that the non-locality is crucial for the protection of the topological qubit: due to the existence of the single-particle gap, there is no way to gap out this zero-energy mode. The only mass term existing at quadratic level is proportial to $i\hat{\gamma}_1\hat{\gamma}_2$, which necessarily involves tunneling between two well-separated points.

\section{Conclusion and Discussion}\label{sec:conclusion}
We study quantum coherence of the Majorana-based topological qubits. We analyze the non-Abelian braiding in the presence of midgap states, and demonstrate that when formulating in terms of the physical observable (fermion parity of the qubit), the braiding statistics is insensitive to the thermal occupation of the midgap states. We also clarify here the conditions for such topological  protection to hold. Our conclusion applies to the case of localized midgap states in the vortex core which are transported along with the Majorana zero states during the braiding operations.
If there are spurious (e.g. impurity-induced~\cite{Lutchyn_disorder, Lutchyn_PRL2011b, Tudor_PRB2011}) midgap bound states spatially located near the Majorana zero-energy states but are not transported together with them, they could strongly affect braiding operations. For example, during braiding the fermion in the qubit has some probability (roughly determined by the non-adiabaticity of the braiding operation) to hybridize with the other bound states near its path leading to an error. If the disorder is weak and short-ranged, such low-energy states are unlikely to occur unless the bulk superconducting gap is significantly suppressed at some spatial points (e.g. vortices) as it is well-known that for a single short-range impurity the energy of such a bound state is close to the bulk excitation gap~\cite{Balatsky_RMP2006, Wang_PRB2004}. Thus, well-separated impurity-induced bound states are typically close to the gap edge and would not affect braiding operations. If the concentration of impurities is increased, then it is meaningful to discuss the probability distribution of the lowest excited bound state in the system~\cite{brouwer_distribution'11}. The distribution of the first excited states determing the minigaps depends on many microscopic details (e.g. system size, concentration of the disorder). Since the magnitude of the minigaps is system-specific, one should evaluate the minigap for a given sample. As a general guiding principle, it is important to reduce the effect of the disorder which limits the speed of braiding operations. However, we note here that physically moving anyons for braiding operations might not be necessary and there are alternative measurement-only approaches to topological quantum computation~\cite{measurement-only} where the issue of the low-lying localized bound states is not relevant.

In this paper we also consider the read-out of topological qubits via interferometry experiments. We study the Mach-Zehnder interferometer based on Aharonov-Casher effect and show that the main effect of midgap states in the Josephson vortices is the reduction of the visibility of the read-out signal. We also consider the effect of thermal excitations involving midgap states of Abrikosov vortices localized in the bulk on the interferometry and find that such processes do not effect the signal provided the system reaches equilibrium fast enough compared to the tunneling time of the Josephson vortices.

Finally, we address the issue of the quantum coherence of the topological qubit itself coupled to a gapped fermionic bath via quantum fluctuations. We derive the master equation governing the time evolution of the reduced density matrix of the topological qubit using a simple physical model Hamiltonian. The decoherence rate of the qubit is exponentially suppressed at low temperatures $T\ll \Delta$. Since topological protection assumes that fermion parity in the superconductor is preserved, our result is very intuitive.

We conclude that the Majorana-based qubits are indeed topologically well-protected  at low temperatures as long as the experimental temperature regime is well below the superconducting gap energy.

\section{Acknowledgement}
We thanks Chang-Yu Hou, Liang Jiang, Chetan Nayak and Kirill Shtengel for illuminating discussion. RL would like to thank the Aspen Center for Physics for hospitality and support under the NSF grant \#1066293. MC would like to thank NSF PHY05-51164 at KITP, UCSB for support as a Graduate Fellow. The work at Maryland (by MC and SDS) is supported by Microsoft Q and DARPA QuEST.

\appendix
\section{Derivation of the Master Equation}\label{appendixA}
In this appendix we derive the master equation for the reduced density matrix.  The density matrix of the whole system evolves according to the equation of motion:
\begin{equation}
	\frac{\di \hat{\rho}}{\di t}=i[\hat{H}_I,\hat{\rho}].
	\label{}
\end{equation}
Notice that we will be working in interaction picture in the following. Here the coupling Hamiltonian
\begin{equation}
	\hat{H}_I(t)=i\sum_{kl}g_{kl}\hat{\gamma}\hat{\eta}_k(t)\hat{\phi}_l(t).
	\label{}
\end{equation}
where
\begin{equation}
\begin{gathered}
	\hat{\eta}_k(t)=\hat{d}_k e^{i\varepsilon_k t}+\hat{d}_k e^{-i\varepsilon_k t},\\
	\hat{\phi}_l(t)=\hat{a}_le^{i\omega_l t}+\hat{a} e^{-i\omega_l t}.
	\label{}
\end{gathered}
\end{equation}
Assume the coupling between the qubit and the bath is weak, we integrate the equation of motion for a time interval $\Delta t$:
\begin{equation}
	\frac{\Delta\hat{\rho}_\text{r}}{\Delta t}=-\frac{1}{\Delta t}\int_t^{t+\Delta t}\di t_1\int_t^{t_1}\di t_2\,\mathrm{Tr}_\mathrm{B}[\hat{{H}}_I(t_1),[\hat{{H}}_I(t_2), \hat{\rho}(t_2)]].
	\label{eqn:me2}
\end{equation}
The first-order term vanishes due to the fact that $\langle \hat{\phi}(t)\rangle=\langle \hat{\eta}(t)\rangle=0$.
Now we make the Born approximation for the bath: assume that the bath is so large that it relaxes very quickly to thermal equilibrium. The density matrix of the whole system can be factorized as $\hat{\rho}(t)=\hat{\rho}_\text{r}(t)\otimes\hat{\rho}_B$. Here the bath includes with the gapped fermionic bath and the bosonic bath.

The commutator on the right-hand side of \eqref{eqn:me2} can be evaluated:
\begin{widetext}
\begin{equation}
	\begin{split}
	\mathrm{Tr}_\mathrm{B}[\hat{{H}}_I(t_1),[\hat{{H}}_I(t_2), \hat{\rho}(t_2)]]\approx
	 \big[\hat{\rho}_\text{r}(t)-\hat{\gamma}(-1)^{\hat{n}}\hat{\rho}_\text{r}(t)(-1)^{\hat{n}}\hat{\gamma}\big]\Big\{\langle\hat{\eta}_k(t_1)\hat{\eta}_k(t_2)\rangle\langle\hat{\phi}_l(t_1)\hat{\phi}_l(t_2)\rangle]+\langle\hat{\eta}_k(t_2)\hat{\eta}_k(t_1)\rangle\langle\hat{\phi}_l(t_2)\hat{\phi}_l(t_1)\rangle]\Big\}
\end{split}
	\label{}
\end{equation}
\end{widetext}
The factor $(-1)^{\hat{n}}$ appears because of the anti-commutation relation between fermionic operators. The correlators of the bath are easily calculated:
\begin{equation}
	\begin{gathered}
	\langle \hat{\eta}_k(t_1)\hat{\eta}_k(t_2)\rangle=n^f_k e^{i\varepsilon_k(t_1-t_2)}+(1-n^f_k)e^{-i\varepsilon_k(t_1-t_2)}\\
	\langle \hat{\phi}_l(t_1)\hat{\phi}_l(t_2)\rangle=n^b_l e^{i\omega_l(t_1-t_2)}+(n^b_l+1)e^{-i\omega_l(t_1-t_2)}
\end{gathered}
	\label{}
\end{equation}

Performing the integral over $t_1$ and $t_2$, we finally arrive at
\begin{equation}
	\frac{\di\hat{\rho}_\text{r}}{\di t}=-\lambda\big[\hat{\rho}_\text{r}-\hat{\gamma}_1(-1)^{\hat{n}}\hat{\rho}_\text{r}(t)(-1)^{\hat{n}}\hat{\gamma}_1\big].
	\label{}
\end{equation}
Here $\lambda$ is given by
\begin{equation}
	 \lambda=2\sum_{kl}g_{kl}^2\big[(1-n^f_{k})n^b_{l}+n^f_{k}(n^b_{l}+1)]\delta(\varepsilon_k-\omega_l).
	\label{}
\end{equation}

\section{Geometric Phases Generated by Midgap Fermions}\label{appendixB}
In this Appendix we derive the geometric phase generated by the midgap fermions, relevant to the interferometry experiments involving Josephson vortices (fluxons). We follow here the formalism developed in the context of AC effect for flux qubits~\cite{Sau_PRA2010}.

We assume that a superconducting island with several midgap fermionic states, labeled by $\hat{d}_m^\dag$, is coupled to a flux qubit. In the low-energy regime well below the bulk superconducting gap and the plasma frequency, the only degrees of freedom of this system are the superconducting phase $\phi$ and the midgap fermions. We also assume that the phase varies slowly so the fermionic part of the system follows the BCS mean-field Hamiltonian with superconducting phase $\phi$.

We want to know the geometric phase associated with vortex tunneling in the presence of midgap fermions. It can be derived by calculating the transition amplitude $\mathcal{A}_{fi}$ associated with a time-depedent phase $\phi=\phi(t)$:
\begin{equation}
	\mathcal{A}_{fi}=\tpd{\phi_f}{\hat{Q}_f \hat{U}(t_f, t_i)\hat{Q}_i^\dag}{\phi_i},
	\label{}
\end{equation}
where $\ket{\phi}$ denotes the BCS ground state with superconducting phase $\phi$ and  $\phi_f-\phi_i=2w\pi$. $\hat{Q}^\dag=\prod_m (\hat{d}_m^\dag)^{n_m}$ denote the occupation of the midgap fermionic states with $n_m=0, 1$.

The midgap fermionic operators $\hat{d}_m^\dag$ are explicity expressed in terms of Bogoliubov wavefunctions $u_m$ and $v_m$:
\begin{equation}
	\hat{d}_m^\dag(t)=e^{-i\varepsilon_m t}\int\di\vr\, \big[u_m(\vr)\hat{\psi}^\dag(\vr)e^{i\phi/2}+v_m(\vr)\hat{\psi}(\vr)e^{-i\phi/2}\big].
	\label{}
\end{equation}
Therefore,
\begin{equation}
	\hat{U}(t_f, t_i)\hat{d}_m^\dag(t_i)\hat{U}^\dag(t_f, t_i)=\hat{d}_m^\dag(t_f) e^{i\pi w {n}_m}.
	\label{}
\end{equation}
So the transition amplitude is evaluated as
\begin{equation}
	\begin{split}	
		\mathcal{A}_{fi}&=\tpd{\phi_f}{\hat{Q}_f \hat{U}(t_f, t_i)\hat{Q}_i^\dag \hat{U}^\dag(t_f,t_i) \hat{U}(t_f, t_i)}{\phi_i}\\
		&=e^{i\pi wn}e^{-i\sum_m n_m\varepsilon_m (t_f-t_i)}\tpd{\phi_f}{\hat{Q}_f\hat{Q}_f^\dag \hat{U}(t_f,t_i)}{\phi_i}\\
		&=e^{i\pi wn}e^{-i\sum_m n_m\varepsilon_m (t_f-t_i)}\tpd{\phi_f}{\hat{U}(t_f, t_i)}{\phi_i}
	\end{split}
		\label{}
\end{equation}
We conclude that the geometric phase is precisely $\pi w n=\frac{n}{2}(\phi_f-\phi_i)$  Physically this reflects the fact that one fermion is ``half'' of a Cooper pair. The vortex tunneling causes the phase of the Cooper pair condensate changes by $2\pi$ and correspondingly the fermionic states obtain $\pi$ phases. Notice that the phase $\sum_m \varepsilon_m(t_f-t_i)$ is simply the overall dynamical phase of the whole system due to its finite energy and does not contribute to the interference at all.
%\bibliography{./refs_braiding}

\begin{thebibliography}{52}
\expandafter\ifx\csname natexlab\endcsname\relax\def\natexlab#1{#1}\fi
\expandafter\ifx\csname bibnamefont\endcsname\relax
  \def\bibnamefont#1{#1}\fi
\expandafter\ifx\csname bibfnamefont\endcsname\relax
  \def\bibfnamefont#1{#1}\fi
\expandafter\ifx\csname citenamefont\endcsname\relax
  \def\citenamefont#1{#1}\fi
\expandafter\ifx\csname url\endcsname\relax
  \def\url#1{\texttt{#1}}\fi
\expandafter\ifx\csname urlprefix\endcsname\relax\def\urlprefix{URL }\fi
\providecommand{\bibinfo}[2]{#2}
\providecommand{\eprint}[2][]{\url{#2}}

\bibitem[{\citenamefont{Kitaev}(2003)}]{Kitaev_AP03}
\bibinfo{author}{\bibfnamefont{A.~Y.} \bibnamefont{Kitaev}},
  \bibinfo{journal}{Ann. Phys.(N.Y.)} \textbf{\bibinfo{volume}{303}},
  \bibinfo{pages}{2 } (\bibinfo{year}{2003}).

\bibitem[{\citenamefont{Nayak et~al.}(2008)\citenamefont{Nayak, Simon, Stern,
  Freedman, and Das~Sarma}}]{nayak_RevModPhys'08}
\bibinfo{author}{\bibfnamefont{C.}~\bibnamefont{Nayak}},
  \bibinfo{author}{\bibfnamefont{S.~H.} \bibnamefont{Simon}},
  \bibinfo{author}{\bibfnamefont{A.}~\bibnamefont{Stern}},
  \bibinfo{author}{\bibfnamefont{M.}~\bibnamefont{Freedman}}, \bibnamefont{and}
  \bibinfo{author}{\bibfnamefont{S.}~\bibnamefont{Das~Sarma}},
  \bibinfo{journal}{Rev. Mod. Phys.} \textbf{\bibinfo{volume}{80}},
  \bibinfo{pages}{1083} (\bibinfo{year}{2008}).

\bibitem[{\citenamefont{{Bonderson} et~al.}(2010)\citenamefont{{Bonderson},
  {Das Sarma}, {Freedman}, and {Nayak}}}]{blueprint}
\bibinfo{author}{\bibfnamefont{P.}~\bibnamefont{{Bonderson}}},
  \bibinfo{author}{\bibfnamefont{S.}~\bibnamefont{{Das Sarma}}},
  \bibinfo{author}{\bibfnamefont{M.}~\bibnamefont{{Freedman}}},
  \bibnamefont{and} \bibinfo{author}{\bibfnamefont{C.}~\bibnamefont{{Nayak}}},
  \bibinfo{journal}{arXiv:1003.2826}  (\bibinfo{year}{2010}).

\bibitem[{\citenamefont{Moore and Read}(1991)}]{Moore_NPB91}
\bibinfo{author}{\bibfnamefont{G.}~\bibnamefont{Moore}} \bibnamefont{and}
  \bibinfo{author}{\bibfnamefont{N.}~\bibnamefont{Read}},
  \bibinfo{journal}{Nucl. Phys. B} \textbf{\bibinfo{volume}{360}},
  \bibinfo{pages}{362 } (\bibinfo{year}{1991}).

\bibitem[{\citenamefont{Nayak and Wilczek}(1996)}]{Nayak_NPB96}
\bibinfo{author}{\bibfnamefont{C.}~\bibnamefont{Nayak}} \bibnamefont{and}
  \bibinfo{author}{\bibfnamefont{F.}~\bibnamefont{Wilczek}},
  \bibinfo{journal}{Nucl. Phys. B} \textbf{\bibinfo{volume}{479}},
  \bibinfo{pages}{529 } (\bibinfo{year}{1996}).

\bibitem[{\citenamefont{Das~Sarma
  et~al.}(2005{\natexlab{a}})\citenamefont{Das~Sarma, Freedman, and
  Nayak}}]{dassarma_prl'05}
\bibinfo{author}{\bibfnamefont{S.}~\bibnamefont{Das~Sarma}},
  \bibinfo{author}{\bibfnamefont{M.}~\bibnamefont{Freedman}}, \bibnamefont{and}
  \bibinfo{author}{\bibfnamefont{C.}~\bibnamefont{Nayak}},
  \bibinfo{journal}{Phys.\ Rev.\ Lett.} \textbf{\bibinfo{volume}{94}},
  \bibinfo{pages}{166802} (\bibinfo{year}{2005}{\natexlab{a}}).

\bibitem[{\citenamefont{Read and Green}(2000)}]{read_prb'00}
\bibinfo{author}{\bibfnamefont{N.}~\bibnamefont{Read}} \bibnamefont{and}
  \bibinfo{author}{\bibfnamefont{D.}~\bibnamefont{Green}},
  \bibinfo{journal}{Phys. Rev. B} \textbf{\bibinfo{volume}{61}},
  \bibinfo{pages}{10267} (\bibinfo{year}{2000}).

\bibitem[{\citenamefont{Kitaev}(2001)}]{Kitaev_Majorana}
\bibinfo{author}{\bibfnamefont{A.}~\bibnamefont{Kitaev}},
  \bibinfo{journal}{Physics-Uspekhi} \textbf{\bibinfo{volume}{44}},
  \bibinfo{pages}{131} (\bibinfo{year}{2001}).

\bibitem[{\citenamefont{Ivanov}(2001)}]{Ivanov_PRL'01}
\bibinfo{author}{\bibfnamefont{D.~A.} \bibnamefont{Ivanov}},
  \bibinfo{journal}{Phys. Rev. Lett.} \textbf{\bibinfo{volume}{86}},
  \bibinfo{pages}{268} (\bibinfo{year}{2001}).

\bibitem[{\citenamefont{Wilczek}(2009)}]{Wilczek_NatPhys}
\bibinfo{author}{\bibfnamefont{F.}~\bibnamefont{Wilczek}},
  \bibinfo{journal}{Nat. Phys.} \textbf{\bibinfo{volume}{5}},
  \bibinfo{pages}{614} (\bibinfo{year}{2009}).

\bibitem[{\citenamefont{Stern}(2010)}]{Stern_Nat}
\bibinfo{author}{\bibfnamefont{A.}~\bibnamefont{Stern}},
  \bibinfo{journal}{Nature} \textbf{\bibinfo{volume}{464}},
  \bibinfo{pages}{187} (\bibinfo{year}{2010}).

\bibitem[{\citenamefont{Franz}(2010)}]{Franz_popular}
\bibinfo{author}{\bibfnamefont{M.}~\bibnamefont{Franz}},
  \bibinfo{journal}{Physics} \textbf{\bibinfo{volume}{3}}, \bibinfo{pages}{24}
  (\bibinfo{year}{2010}).

\bibitem[{\citenamefont{Levi}(2011)}]{Levi_PhysToday}
\bibinfo{author}{\bibfnamefont{B.~G.} \bibnamefont{Levi}},
  \bibinfo{journal}{Physics Today} \textbf{\bibinfo{volume}{64}},
  \bibinfo{pages}{20} (\bibinfo{year}{2011}).

\bibitem[{\citenamefont{Das~Sarma et~al.}(2006)\citenamefont{Das~Sarma, Nayak,
  and Tewari}}]{DasSarma_PRB'06}
\bibinfo{author}{\bibfnamefont{S.}~\bibnamefont{Das~Sarma}},
  \bibinfo{author}{\bibfnamefont{C.}~\bibnamefont{Nayak}}, \bibnamefont{and}
  \bibinfo{author}{\bibfnamefont{S.}~\bibnamefont{Tewari}},
  \bibinfo{journal}{Phys.\ Rev.\ B} \textbf{\bibinfo{volume}{73}},
  \bibinfo{pages}{220502} (\bibinfo{year}{2006}).

\bibitem[{\citenamefont{Fu and Kane}(2008)}]{Fu_PRL08}
\bibinfo{author}{\bibfnamefont{L.}~\bibnamefont{Fu}} \bibnamefont{and}
  \bibinfo{author}{\bibfnamefont{C.~L.} \bibnamefont{Kane}},
  \bibinfo{journal}{Phys. Rev. Lett.} \textbf{\bibinfo{volume}{100}},
  \bibinfo{pages}{096407} (\bibinfo{year}{2008}).

\bibitem[{\citenamefont{Cook and Franz}(2011)}]{Cook_PRB2011}
\bibinfo{author}{\bibfnamefont{A.}~\bibnamefont{Cook}} \bibnamefont{and}
  \bibinfo{author}{\bibfnamefont{M.}~\bibnamefont{Franz}},
  \bibinfo{journal}{Phys. Rev. B} \textbf{\bibinfo{volume}{84}},
  \bibinfo{pages}{201105} (\bibinfo{year}{2011}).

\bibitem[{\citenamefont{Sau et~al.}(2010{\natexlab{a}})\citenamefont{Sau,
  Lutchyn, Tewari, and Das~Sarma}}]{Sau_PRL10}
\bibinfo{author}{\bibfnamefont{J.~D.} \bibnamefont{Sau}},
  \bibinfo{author}{\bibfnamefont{R.~M.} \bibnamefont{Lutchyn}},
  \bibinfo{author}{\bibfnamefont{S.}~\bibnamefont{Tewari}}, \bibnamefont{and}
  \bibinfo{author}{\bibfnamefont{S.}~\bibnamefont{Das~Sarma}},
  \bibinfo{journal}{Phys. Rev. Lett.} \textbf{\bibinfo{volume}{104}},
  \bibinfo{pages}{040502} (\bibinfo{year}{2010}{\natexlab{a}}).

\bibitem[{\citenamefont{Alicea}(2010)}]{Alicea_PRB10}
\bibinfo{author}{\bibfnamefont{J.}~\bibnamefont{Alicea}},
  \bibinfo{journal}{Phys. Rev. B} \textbf{\bibinfo{volume}{81}},
  \bibinfo{pages}{125318} (\bibinfo{year}{2010}).

\bibitem[{\citenamefont{Lutchyn et~al.}(2010)\citenamefont{Lutchyn, Sau, and
  Das~Sarma}}]{Lutchyn_PRL2011}
\bibinfo{author}{\bibfnamefont{R.~M.} \bibnamefont{Lutchyn}},
  \bibinfo{author}{\bibfnamefont{J.~D.} \bibnamefont{Sau}}, \bibnamefont{and}
  \bibinfo{author}{\bibfnamefont{S.}~\bibnamefont{Das~Sarma}},
  \bibinfo{journal}{Phys. Rev. Lett.} \textbf{\bibinfo{volume}{105}},
  \bibinfo{pages}{077001} (\bibinfo{year}{2010}).

\bibitem[{\citenamefont{Oreg et~al.}(2010)\citenamefont{Oreg, Refael, and von
  Oppen}}]{Oreg_PRL2010}
\bibinfo{author}{\bibfnamefont{Y.}~\bibnamefont{Oreg}},
  \bibinfo{author}{\bibfnamefont{G.}~\bibnamefont{Refael}}, \bibnamefont{and}
  \bibinfo{author}{\bibfnamefont{F.}~\bibnamefont{von Oppen}},
  \bibinfo{journal}{Phys. Rev. Lett.} \textbf{\bibinfo{volume}{105}},
  \bibinfo{pages}{177002} (\bibinfo{year}{2010}).

\bibitem[{\citenamefont{Qi et~al.}(2010)\citenamefont{Qi, Hughes, and
  Zhang}}]{Qi_PRB2010}
\bibinfo{author}{\bibfnamefont{X.-L.} \bibnamefont{Qi}},
  \bibinfo{author}{\bibfnamefont{T.~L.} \bibnamefont{Hughes}},
  \bibnamefont{and} \bibinfo{author}{\bibfnamefont{S.-C.} \bibnamefont{Zhang}},
  \bibinfo{journal}{Phys. Rev. B} \textbf{\bibinfo{volume}{82}},
  \bibinfo{pages}{184516} (\bibinfo{year}{2010}).

\bibitem[{\citenamefont{Sato and Fujimoto}(2009)}]{Sato_PRB09}
\bibinfo{author}{\bibfnamefont{M.}~\bibnamefont{Sato}} \bibnamefont{and}
  \bibinfo{author}{\bibfnamefont{S.}~\bibnamefont{Fujimoto}},
  \bibinfo{journal}{Phys. Rev. B} \textbf{\bibinfo{volume}{79}},
  \bibinfo{pages}{094504} (\bibinfo{year}{2009}).

\bibitem[{\citenamefont{Bonderson et~al.}(2011)\citenamefont{Bonderson,
  Gurarie, and Nayak}}]{Bonderson_PRB2011}
\bibinfo{author}{\bibfnamefont{P.}~\bibnamefont{Bonderson}},
  \bibinfo{author}{\bibfnamefont{V.}~\bibnamefont{Gurarie}}, \bibnamefont{and}
  \bibinfo{author}{\bibfnamefont{C.}~\bibnamefont{Nayak}},
  \bibinfo{journal}{Phys. Rev. B} \textbf{\bibinfo{volume}{83}},
  \bibinfo{pages}{075303} (\bibinfo{year}{2011}).

\bibitem[{\citenamefont{Alicea et~al.}(2011)\citenamefont{Alicea, Oreg, Refael,
  von Oppen, and Fisher}}]{Alicea_NatPhys2011}
\bibinfo{author}{\bibfnamefont{J.}~\bibnamefont{Alicea}},
  \bibinfo{author}{\bibfnamefont{Y.}~\bibnamefont{Oreg}},
  \bibinfo{author}{\bibfnamefont{G.}~\bibnamefont{Refael}},
  \bibinfo{author}{\bibfnamefont{F.}~\bibnamefont{von Oppen}},
  \bibnamefont{and} \bibinfo{author}{\bibfnamefont{M.~P.~A.}
  \bibnamefont{Fisher}}, \bibinfo{journal}{Nat. Phys.}
  \textbf{\bibinfo{volume}{7}}, \bibinfo{pages}{412} (\bibinfo{year}{2011}).

\bibitem[{\citenamefont{Cheng et~al.}(2009)\citenamefont{Cheng, Lutchyn,
  Galitski, and Das~Sarma}}]{Cheng_PRL09}
\bibinfo{author}{\bibfnamefont{M.}~\bibnamefont{Cheng}},
  \bibinfo{author}{\bibfnamefont{R.~M.} \bibnamefont{Lutchyn}},
  \bibinfo{author}{\bibfnamefont{V.}~\bibnamefont{Galitski}}, \bibnamefont{and}
  \bibinfo{author}{\bibfnamefont{S.}~\bibnamefont{Das~Sarma}},
  \bibinfo{journal}{Phys. Rev. Lett.} \textbf{\bibinfo{volume}{103}},
  \bibinfo{pages}{107001} (\bibinfo{year}{2009}).

\bibitem[{\citenamefont{Cheng et~al.}(2010)\citenamefont{Cheng, Lutchyn,
  Galitski, and Das~Sarma}}]{Cheng_PRB2010b}
\bibinfo{author}{\bibfnamefont{M.}~\bibnamefont{Cheng}},
  \bibinfo{author}{\bibfnamefont{R.~M.} \bibnamefont{Lutchyn}},
  \bibinfo{author}{\bibfnamefont{V.}~\bibnamefont{Galitski}}, \bibnamefont{and}
  \bibinfo{author}{\bibfnamefont{S.}~\bibnamefont{Das~Sarma}},
  \bibinfo{journal}{Phys. Rev. B} \textbf{\bibinfo{volume}{82}},
  \bibinfo{pages}{094504} (\bibinfo{year}{2010}).

\bibitem[{\citenamefont{Sau et~al.}(2011{\natexlab{a}})\citenamefont{Sau, Lin,
  Hui, and Das~Sarma}}]{Sau_periodic}
\bibinfo{author}{\bibfnamefont{J.~D.} \bibnamefont{Sau}},
  \bibinfo{author}{\bibfnamefont{C.~H.} \bibnamefont{Lin}},
  \bibinfo{author}{\bibfnamefont{H.-Y.} \bibnamefont{Hui}}, \bibnamefont{and}
  \bibinfo{author}{\bibfnamefont{S.}~\bibnamefont{Das~Sarma}},
  \bibinfo{journal}{arXiv:1103.2770}  (\bibinfo{year}{2011}{\natexlab{a}}).

\bibitem[{\citenamefont{Caroli et~al.}(1964)\citenamefont{Caroli, de~Gennes,
  and Matricon}}]{Caroli_PL'64}
\bibinfo{author}{\bibfnamefont{C.}~\bibnamefont{Caroli}},
  \bibinfo{author}{\bibfnamefont{P.}~\bibnamefont{de~Gennes}},
  \bibnamefont{and} \bibinfo{author}{\bibfnamefont{J.}~\bibnamefont{Matricon}},
  \bibinfo{journal}{Phys. Lett.} \textbf{\bibinfo{volume}{9}},
  \bibinfo{pages}{307} (\bibinfo{year}{1964}).

\bibitem[{\citenamefont{Kopnin and Salomaa}(1991)}]{Kopnin_PRB'91}
\bibinfo{author}{\bibfnamefont{N.~B.} \bibnamefont{Kopnin}} \bibnamefont{and}
  \bibinfo{author}{\bibfnamefont{M.~M.} \bibnamefont{Salomaa}},
  \bibinfo{journal}{Phys. Rev. B} \textbf{\bibinfo{volume}{44}},
  \bibinfo{pages}{9667} (\bibinfo{year}{1991}).

\bibitem[{\citenamefont{Sau et~al.}(2010{\natexlab{b}})\citenamefont{Sau,
  Lutchyn, Tewari, and Das~Sarma}}]{robustness}
\bibinfo{author}{\bibfnamefont{J.~D.} \bibnamefont{Sau}},
  \bibinfo{author}{\bibfnamefont{R.~M.} \bibnamefont{Lutchyn}},
  \bibinfo{author}{\bibfnamefont{S.}~\bibnamefont{Tewari}}, \bibnamefont{and}
  \bibinfo{author}{\bibfnamefont{S.}~\bibnamefont{Das~Sarma}},
  \bibinfo{journal}{Phys. Rev. B} \textbf{\bibinfo{volume}{82}},
  \bibinfo{pages}{094522} (\bibinfo{year}{2010}{\natexlab{b}}).

\bibitem[{\citenamefont{Goldstein and Chamon}(2011)}]{Goldstein_wrong}
\bibinfo{author}{\bibfnamefont{G.}~\bibnamefont{Goldstein}} \bibnamefont{and}
  \bibinfo{author}{\bibfnamefont{C.}~\bibnamefont{Chamon}},
  \bibinfo{journal}{arXiv:1107.0288}  (\bibinfo{year}{2011}).

\bibitem[{\citenamefont{Akhmerov}(2010)}]{Akhmerov_PRB2010}
\bibinfo{author}{\bibfnamefont{A.~R.} \bibnamefont{Akhmerov}},
  \bibinfo{journal}{Phys. Rev. B} \textbf{\bibinfo{volume}{82}},
  \bibinfo{pages}{020509} (\bibinfo{year}{2010}).

\bibitem[{\citenamefont{Cheng et~al.}(2011)\citenamefont{Cheng, Galitski, and
  Das~Sarma}}]{Cheng_PRB2011}
\bibinfo{author}{\bibfnamefont{M.}~\bibnamefont{Cheng}},
  \bibinfo{author}{\bibfnamefont{V.}~\bibnamefont{Galitski}}, \bibnamefont{and}
  \bibinfo{author}{\bibfnamefont{S.}~\bibnamefont{Das~Sarma}},
  \bibinfo{journal}{Phys. Rev. B} \textbf{\bibinfo{volume}{84}},
  \bibinfo{pages}{104529} (\bibinfo{year}{2011}).

\bibitem[{\citenamefont{Lindblad}(1976)}]{lindblad}
\bibinfo{author}{\bibfnamefont{G.}~\bibnamefont{Lindblad}},
  \bibinfo{journal}{Commun. Math. Phys.} \textbf{\bibinfo{volume}{48}},
  \bibinfo{pages}{119} (\bibinfo{year}{1976}).

\bibitem[{\citenamefont{Das~Sarma
  et~al.}(2005{\natexlab{b}})\citenamefont{Das~Sarma, Freedman, and
  Nayak}}]{SDS_PRL2005}
\bibinfo{author}{\bibfnamefont{S.}~\bibnamefont{Das~Sarma}},
  \bibinfo{author}{\bibfnamefont{M.}~\bibnamefont{Freedman}}, \bibnamefont{and}
  \bibinfo{author}{\bibfnamefont{C.}~\bibnamefont{Nayak}},
  \bibinfo{journal}{Phys. Rev. Lett.} \textbf{\bibinfo{volume}{94}},
  \bibinfo{pages}{166802} (\bibinfo{year}{2005}{\natexlab{b}}).

\bibitem[{\citenamefont{Stern and Halperin}(2006)}]{Stern_PRL2006}
\bibinfo{author}{\bibfnamefont{A.}~\bibnamefont{Stern}} \bibnamefont{and}
  \bibinfo{author}{\bibfnamefont{B.~I.} \bibnamefont{Halperin}},
  \bibinfo{journal}{Phys. Rev. Lett.} \textbf{\bibinfo{volume}{96}},
  \bibinfo{pages}{016802} (\bibinfo{year}{2006}).

\bibitem[{\citenamefont{Bonderson et~al.}(2006)\citenamefont{Bonderson, Kitaev,
  and Shtengel}}]{Bonderson_PRL2006}
\bibinfo{author}{\bibfnamefont{P.}~\bibnamefont{Bonderson}},
  \bibinfo{author}{\bibfnamefont{A.}~\bibnamefont{Kitaev}}, \bibnamefont{and}
  \bibinfo{author}{\bibfnamefont{K.}~\bibnamefont{Shtengel}},
  \bibinfo{journal}{Phys. Rev. Lett.} \textbf{\bibinfo{volume}{96}},
  \bibinfo{pages}{016803} (\bibinfo{year}{2006}).

\bibitem[{\citenamefont{Bonderson
  et~al.}(2008{\natexlab{a}})\citenamefont{Bonderson, Shtengel, and
  Slingerland}}]{Bonderson_AnnPhys2008}
\bibinfo{author}{\bibfnamefont{P.}~\bibnamefont{Bonderson}},
  \bibinfo{author}{\bibfnamefont{K.}~\bibnamefont{Shtengel}}, \bibnamefont{and}
  \bibinfo{author}{\bibfnamefont{J.}~\bibnamefont{Slingerland}},
  \bibinfo{journal}{Ann. Phys. (N. Y.)} \textbf{\bibinfo{volume}{323}},
  \bibinfo{pages}{2709} (\bibinfo{year}{2008}{\natexlab{a}}), ISSN
  \bibinfo{issn}{0003-4916}.

\bibitem[{\citenamefont{Akhmerov et~al.}(2009)\citenamefont{Akhmerov, Nilsson,
  and Beenakker}}]{Akhmerov_int'09}
\bibinfo{author}{\bibfnamefont{A.~R.} \bibnamefont{Akhmerov}},
  \bibinfo{author}{\bibfnamefont{J.}~\bibnamefont{Nilsson}}, \bibnamefont{and}
  \bibinfo{author}{\bibfnamefont{C.~W.~J.} \bibnamefont{Beenakker}},
  \bibinfo{journal}{Phys. Rev. Lett.} \textbf{\bibinfo{volume}{102}},
  \bibinfo{pages}{216404} (\bibinfo{year}{2009}).

\bibitem[{\citenamefont{Sau et~al.}(2011{\natexlab{b}})\citenamefont{Sau,
  Tewari, and Das~Sarma}}]{Sau_arxiv1004}
\bibinfo{author}{\bibfnamefont{J.~D.} \bibnamefont{Sau}},
  \bibinfo{author}{\bibfnamefont{S.}~\bibnamefont{Tewari}}, \bibnamefont{and}
  \bibinfo{author}{\bibfnamefont{S.}~\bibnamefont{Das~Sarma}},
  \bibinfo{journal}{Phys. Rev. B} \textbf{\bibinfo{volume}{84}},
  \bibinfo{pages}{085109} (\bibinfo{year}{2011}{\natexlab{b}}).

\bibitem[{\citenamefont{Grosfeld et~al.}(2011)\citenamefont{Grosfeld, Seradjeh,
  and Vishveshwara}}]{Grosfeld_PRB2010}
\bibinfo{author}{\bibfnamefont{E.}~\bibnamefont{Grosfeld}},
  \bibinfo{author}{\bibfnamefont{B.}~\bibnamefont{Seradjeh}}, \bibnamefont{and}
  \bibinfo{author}{\bibfnamefont{S.}~\bibnamefont{Vishveshwara}},
  \bibinfo{journal}{Phys. Rev. B} \textbf{\bibinfo{volume}{83}},
  \bibinfo{pages}{104513} (\bibinfo{year}{2011}).

\bibitem[{\citenamefont{Grosfeld and Stern}(2011)}]{Grosfeld_PNAS}
\bibinfo{author}{\bibfnamefont{E.}~\bibnamefont{Grosfeld}} \bibnamefont{and}
  \bibinfo{author}{\bibfnamefont{A.}~\bibnamefont{Stern}},
  \bibinfo{journal}{Proc. Natl. Acad. Sci. USA} \textbf{\bibinfo{volume}{108}},
  \bibinfo{pages}{11810} (\bibinfo{year}{2011}).

\bibitem[{\citenamefont{Sau et~al.}(2010{\natexlab{c}})\citenamefont{Sau,
  Tewari, and Das~Sarma}}]{Sau_PRA2010}
\bibinfo{author}{\bibfnamefont{J.~D.} \bibnamefont{Sau}},
  \bibinfo{author}{\bibfnamefont{S.}~\bibnamefont{Tewari}}, \bibnamefont{and}
  \bibinfo{author}{\bibfnamefont{S.}~\bibnamefont{Das~Sarma}},
  \bibinfo{journal}{Phys. Rev. A} \textbf{\bibinfo{volume}{82}},
  \bibinfo{pages}{052322} (\bibinfo{year}{2010}{\natexlab{c}}).

\bibitem[{\citenamefont{Hassler et~al.}(2010)\citenamefont{Hassler, Akhmerov,
  Hou, and Beenakker}}]{Hassler_NJP2010}
\bibinfo{author}{\bibfnamefont{F.}~\bibnamefont{Hassler}},
  \bibinfo{author}{\bibfnamefont{A.~R.} \bibnamefont{Akhmerov}},
  \bibinfo{author}{\bibfnamefont{C.-Y.} \bibnamefont{Hou}}, \bibnamefont{and}
  \bibinfo{author}{\bibfnamefont{C.~W.~J.} \bibnamefont{Beenakker}},
  \bibinfo{journal}{New J. Phys.} \textbf{\bibinfo{volume}{12}},
  \bibinfo{pages}{125002} (\bibinfo{year}{2010}).

\bibitem[{\citenamefont{Clarke and Shtengel}(2011)}]{Clarke_NJP2011}
\bibinfo{author}{\bibfnamefont{D.~J.} \bibnamefont{Clarke}} \bibnamefont{and}
  \bibinfo{author}{\bibfnamefont{K.}~\bibnamefont{Shtengel}},
  \bibinfo{journal}{New J. Phys.} \textbf{\bibinfo{volume}{13}},
  \bibinfo{pages}{055005} (\bibinfo{year}{2011}).

\bibitem[{\citenamefont{Lutchyn
  et~al.}(2011{\natexlab{a}})\citenamefont{Lutchyn, Stanescu, and
  Das~Sarma}}]{Lutchyn_disorder}
\bibinfo{author}{\bibfnamefont{R.~M.} \bibnamefont{Lutchyn}},
  \bibinfo{author}{\bibfnamefont{T.~D.} \bibnamefont{Stanescu}},
  \bibnamefont{and}
  \bibinfo{author}{\bibfnamefont{S.}~\bibnamefont{Das~Sarma}},
  \bibinfo{journal}{arXiv:1011.5643}  (\bibinfo{year}{2011}{\natexlab{a}}).

\bibitem[{\citenamefont{Lutchyn
  et~al.}(2011{\natexlab{b}})\citenamefont{Lutchyn, Stanescu, and
  Das~Sarma}}]{Lutchyn_PRL2011b}
\bibinfo{author}{\bibfnamefont{R.~M.} \bibnamefont{Lutchyn}},
  \bibinfo{author}{\bibfnamefont{T.~D.} \bibnamefont{Stanescu}},
  \bibnamefont{and}
  \bibinfo{author}{\bibfnamefont{S.}~\bibnamefont{Das~Sarma}},
  \bibinfo{journal}{Phys. Rev. Lett.} \textbf{\bibinfo{volume}{106}},
  \bibinfo{pages}{127001} (\bibinfo{year}{2011}{\natexlab{b}}).

\bibitem[{\citenamefont{Stanescu et~al.}(2011)\citenamefont{Stanescu, Lutchyn,
  and Das~Sarma}}]{Tudor_PRB2011}
\bibinfo{author}{\bibfnamefont{T.~D.} \bibnamefont{Stanescu}},
  \bibinfo{author}{\bibfnamefont{R.~M.} \bibnamefont{Lutchyn}},
  \bibnamefont{and}
  \bibinfo{author}{\bibfnamefont{S.}~\bibnamefont{Das~Sarma}},
  \bibinfo{journal}{Phys. Rev. B} \textbf{\bibinfo{volume}{84}},
  \bibinfo{pages}{144522} (\bibinfo{year}{2011}).

\bibitem[{\citenamefont{Balatsky et~al.}(2006)\citenamefont{Balatsky, Vekhter,
  and Zhu}}]{Balatsky_RMP2006}
\bibinfo{author}{\bibfnamefont{A.~V.} \bibnamefont{Balatsky}},
  \bibinfo{author}{\bibfnamefont{I.}~\bibnamefont{Vekhter}}, \bibnamefont{and}
  \bibinfo{author}{\bibfnamefont{J.-X.} \bibnamefont{Zhu}},
  \bibinfo{journal}{Rev. Mod. Phys.} \textbf{\bibinfo{volume}{78}},
  \bibinfo{pages}{373} (\bibinfo{year}{2006}).

\bibitem[{\citenamefont{Wang and Wang}(2004)}]{Wang_PRB2004}
\bibinfo{author}{\bibfnamefont{Q.-H.} \bibnamefont{Wang}} \bibnamefont{and}
  \bibinfo{author}{\bibfnamefont{Z.~D.} \bibnamefont{Wang}},
  \bibinfo{journal}{Phys. Rev. B} \textbf{\bibinfo{volume}{69}},
  \bibinfo{pages}{092502} (\bibinfo{year}{2004}).

\bibitem[{\citenamefont{Brouwer et~al.}(2011)\citenamefont{Brouwer, Duckheim,
  Romito, and von Oppen}}]{brouwer_distribution'11}
\bibinfo{author}{\bibfnamefont{P.~W.} \bibnamefont{Brouwer}},
  \bibinfo{author}{\bibfnamefont{M.}~\bibnamefont{Duckheim}},
  \bibinfo{author}{\bibfnamefont{A.}~\bibnamefont{Romito}}, \bibnamefont{and}
  \bibinfo{author}{\bibfnamefont{F.}~\bibnamefont{von Oppen}},
  \bibinfo{journal}{Phys. Rev. Lett.} \textbf{\bibinfo{volume}{107}},
  \bibinfo{pages}{196804} (\bibinfo{year}{2011}).

\bibitem[{\citenamefont{Bonderson
  et~al.}(2008{\natexlab{b}})\citenamefont{Bonderson, Freedman, and
  Nayak}}]{measurement-only}
\bibinfo{author}{\bibfnamefont{P.}~\bibnamefont{Bonderson}},
  \bibinfo{author}{\bibfnamefont{M.}~\bibnamefont{Freedman}}, \bibnamefont{and}
  \bibinfo{author}{\bibfnamefont{C.}~\bibnamefont{Nayak}},
  \bibinfo{journal}{Phys. Rev. Lett.} \textbf{\bibinfo{volume}{101}},
  \bibinfo{pages}{010501} (\bibinfo{year}{2008}{\natexlab{b}}).

\end{thebibliography}

\end{document}